\documentclass[superscriptaddress,
reprint,
 amsmath,amssymb,
 aps,
pra,
]{revtex4-2}

\usepackage{graphicx}
\usepackage{dcolumn}
\usepackage{bm}
\usepackage{hyperref}
\usepackage{lineno}
\usepackage{color,soul}

\usepackage[normalem]{ulem}
\usepackage{soul}
\sethlcolor{yellow} 

\begin{document}

\preprint{APS/123-QED}

\title{Static dc electric field orientation effects on two-photon Rydberg EIT}

\author{Rob Behary}
\email{rbehary@wm.edu}
\affiliation{Department of Physics, William \& Mary, 300 Ukrop Way, Williamsburg, VA 23185, USA}
\author{William Torg}
\affiliation{Department of Physics, William \& Mary, 300 Ukrop Way, Williamsburg, VA 23185, USA}
\author{Mykhailo Vorobiov}
\affiliation{Department of Physics, William \& Mary, 300 Ukrop Way, Williamsburg, VA 23185, USA}
\author{Nicolas DeStefano}
\affiliation{Department of Physics, William \& Mary, 300 Ukrop Way, Williamsburg, VA 23185, USA}
\author{Adam Vernon}
\affiliation{Department of Physics, William \& Mary, 300 Ukrop Way, Williamsburg, VA 23185, USA}
\author{Charles T. Fancher}
\affiliation{The MITRE Corporation, McLean, VA 22102, USA}
\author{Neel Malvania}
\affiliation{The MITRE Corporation, McLean, VA 22102, USA}
\author{Eugeniy E. Mikhailov}
\affiliation{Department of Physics, William \& Mary, 300 Ukrop Way, Williamsburg, VA 23185, USA}
\author{Seth Aubin}
\affiliation{Department of Physics, William \& Mary, 300 Ukrop Way, Williamsburg, VA 23185, USA}
\author{Irina Novikova}
\affiliation{Department of Physics, William \& Mary, 300 Ukrop Way, Williamsburg, VA 23185, USA}

\date{\today}

\begin{abstract}
We examine the influence of a static dc electric field on Electromagnetically Induced Transparency (EIT) resonances that involve highly excited Rydberg states. 
Our focus is on how these resonances are altered when the relative orientation between the laser polarization and the  external electric field vectors are varied. We experimentally demonstrate characteristic variations in the amplitude of the Stark-split EIT resonances, which can be explained by the selection rules in various geometries. We also present a simplified semi-analytical model that closely resembles the experimental observations. 
We use these findings to obtain information about the spatially inhomogeneous electric field, produced by a biased wire, using EIT fluorescence measurements that agrees with the expected angular dependencies. These results suggest that simultaneous analysis of frequency shifts and amplitudes of Rydberg EIT resonances may enable vector electrometry of electrostatic fields, necessary for many quantum sensing applications.
\end{abstract}

\maketitle

\section{Introduction}
Complete characterization of any vector requires information about its magnitude and direction. The large polarizability of alkali metal atoms in highly excited Rydberg states makes them attractive for scalar electric field sensing~\cite{Gallagher_1994}, and in the last two decades many research groups have demonstrated rf and THz field sensors using room temperature Rb or Cs vapor cells~\cite{HollowayJAP2017,Nik_Rydberg_Review,KermitRFReciever,FrancherIEEE2021,AdamsTHzImaging,Downes_2023,PhysRevX.10.011027,RydbergBlackbody}. The majority of such sensors rely on coherent two- or three-photon electromagnetically induced transparency (EIT)~\cite{Finkelstein_2023,Nik_Rydberg_Review,SIMONS2021100273} to detect changes in the Rydberg state energies in the presence of external dc or ac electric fields. However, the quadratic dependence of the Stark shifts only allows information about the electric field magnitude. The directional information may be obtained by interfering the test field with a local oscillator of known polarization~\cite{PhysRevApplied.22.064012}. However, this approach is not practical for measurements of low-frequency or dc electric fields with free charges, as any additional electric field will modify the original charge distribution and thus disturb the electric environment to be measured.

In this manuscript we attempt to reconstruct a dc electric field vector by recording both the frequencies and areas of EIT two-photon resonances for different sub-levels of a Rydberg state. This approach relies on the polarization dependence of the transition probabilities between various Zeeman sub-levels, and has been explored previously to determine the direction of magnetic fields~\cite{Yudin2010,mikhailov2010compass,GonzalezMaldonadoOE24} and rf fields~\cite{PhysRevLett.111.063001,chen2025polarizationawaredoadetectionrelying}. 
In our experiments we demonstrate that we can determine the orientation of the electric field inside a vacuum chamber filled with Rb atoms by rotating the laser polarization and tracking changes in amplitudes and areas of Stark-split EIT peaks. Detection of EIT-induced fluorescence dips allows us to obtain spatial information about the inhomogeneous electric field, and reconstruct changes in its magnitude and orientation. 
We also present a semi-analytical atomic model that qualitatively agrees with our experimental data.

\begin{figure}[h!]
    \centering
    \includegraphics[width=\linewidth]{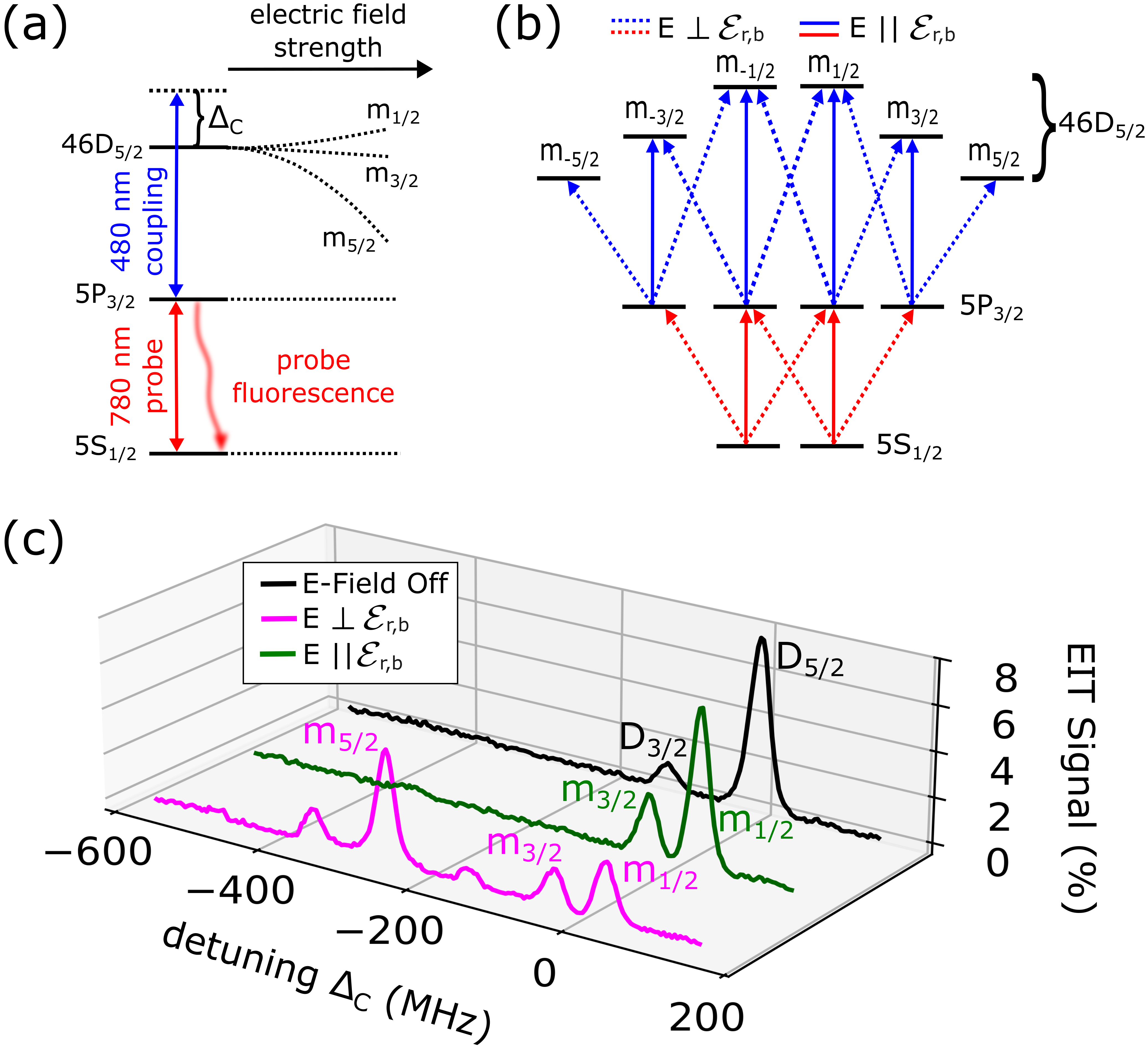}
    \caption{(a) 
        Simplified energy level configuration of $^{85}$Rb used to observe two-photon EIT resonances. 
        The 780~nm probe laser is on resonance with the $5S_{1/2}\rightarrow 5P_{3/2}$ transition. The 480~nm coupling laser is scanned across the $5P_{3/2}\rightarrow nD_{5/2}$ transition with frequency detuning $\Delta_c$. 
        Dashed lines depict Stark splitting of the Rydberg $nD_{5/2}$ level into $|m_J|=1/2,3/2,5/2$ sublevels as the static electric field strength increases. 
        (b) Allowed transitions for optical field polarized parallel (solid) or perpendicular (dashed) to the dc electric field for a simplified fine structure of involved atomic levels.
        (c) Examples of EIT spectra in the absence of electric field (black), with electric field \textbf{E} applied perpendicular (magenta) and parallel (green) to both laser polarizations ($\mathcal{E}$). 
    }
    \label{fig:ladder-scheme}
\end{figure}

\section{Basic Sensing Scheme}
In our experiment we employ a ladder-type EIT scheme to excite Rb atoms to the  46D Rydberg levels using two counter-propagating lasers at 780~nm and 480~nm, as shown in Fig.~\ref{fig:ladder-scheme}(a). 
In the presence of electric fields both fine structure levels of the Rydberg state are split into three ($J=5/2$) or two ($J=3/2$) dc-Stark shifted energy sublevels where the shift strength is given by:
\begin{equation}
\label{eq:stark-shift}
    h\Delta f_{|m_J|}(E) = -\frac{1}{2}\alpha_{|m_J|}E^2.
\end{equation}
Where $h$ is Planck's constant, $\alpha_{|m_J|}$ is the polarizability of the different $|m_J|$ sublevels of the Rydberg state and $E$ is the magnitude of the electric field~\cite{Gallagher_1994, HollowayJAP2017}. 
Every time the frequency sum of the two lasers precisely matches the energy difference between the ground atomic state and one of the Rydberg states, the atoms are promoted into a coherent ``dark'' superposition of these two states, consequently reducing the population of the intermediate, short-lived $5P_{3/2}$ state. This allows us to determine the values of the Stark shifts for all these Rydberg levels using either transmission resonances in the EIT spectrum or corresponding dips in Rb fluorescence at 780~nm~\cite{SchlossbergerOL25,BeharyPRR2025,surfaceCharges,StandingWaveFluorescence,pati2025millimeterwaveimagingusing}. 


The relative orientation of the electric field with respect to the light field propagation direction and polarization does not affect the frequency positions of the EIT resonances, thanks to the quadratic nature of the Stark shift (at least for relatively  low  electric field amplitudes that do not produce Rydberg state mixing). However, the coupling strength to individual Zeeman sublevels of the Rydberg state is fairly polarization sensitive. Selection rules dictate that the laser polarization ($\mathcal{E}$) component parallel to the electric field ($\pi$-polarization) can only excite transitions with $\Delta m = 0$, while the circularly polarized components ($\sigma_\pm$) are responsible correspondingly for $\Delta m = \pm 1$ transitions. As a result, the populations of the various Rydberg sublevels, and thus the amplitudes of the associated EIT resonances for a given electric field orientation are determined by the choice of both polarizations for the red ($\mathcal{E}_r$) and blue ($\mathcal{E}_b$) lasers.
Using a simplified energy level diagram that ignores the hyperfine structure, shown in Fig.~\ref{fig:ladder-scheme}(b), it can be seen that if both lasers are polarized along the electric field, only $\Delta m = 0$ transitions are allowed, and thus the transitions to the Rydberg state with $m_{\pm5/2}$ become impossible. At the same time, the perpendicular linear laser polarizations drive $\Delta m = \pm 1$ effectively maximize these transitions. 
This qualitative analysis supports the experimental observations of EIT spectra shown in Fig.~\ref{fig:ladder-scheme}(c). For these measurements both laser fields have identical linearly polarizations, directed either perpendicular or parallel to the constant electric field.

When the laser polarizations are perpendicular to the electric field, we observe all three Stark-shifted EIT peaks within the ($J=5/2$) Rydberg transition. Indeed, in this case only optical transitions with $\Delta m=\pm 1$ are active, as shown in dashed lines in Fig.~\ref{fig:ladder-scheme}(b). In this situation, the $m_{\pm 5/2}$ EIT peak has the highest amplitude due to the largest transition matrix elements involved, as has been previously reported~\cite{MaOE20}. At the same time the amplitude of $m_{\pm 1/2}$ EIT peak is relatively low.

However, when the lasers are polarized along the electric field, the $m_{\pm 5/2}$ EIT peak completely disappears, since only $\Delta m=0$ transitions are allowed. It is easy to see that in this case there is no excitation path to reach states with $m_{\pm 5/2}$, and consequently this peak is suppressed in the transmission spectrum. 
Thus, we expect that by rotating the polarization and tracking the amplitudes of Rydberg EIT resonances, we should be able to determine the relative contributions of different transitions and thus extract the information about the electric field orientation.

\begin{figure*}
    \centering
    \includegraphics[width=1\textwidth]{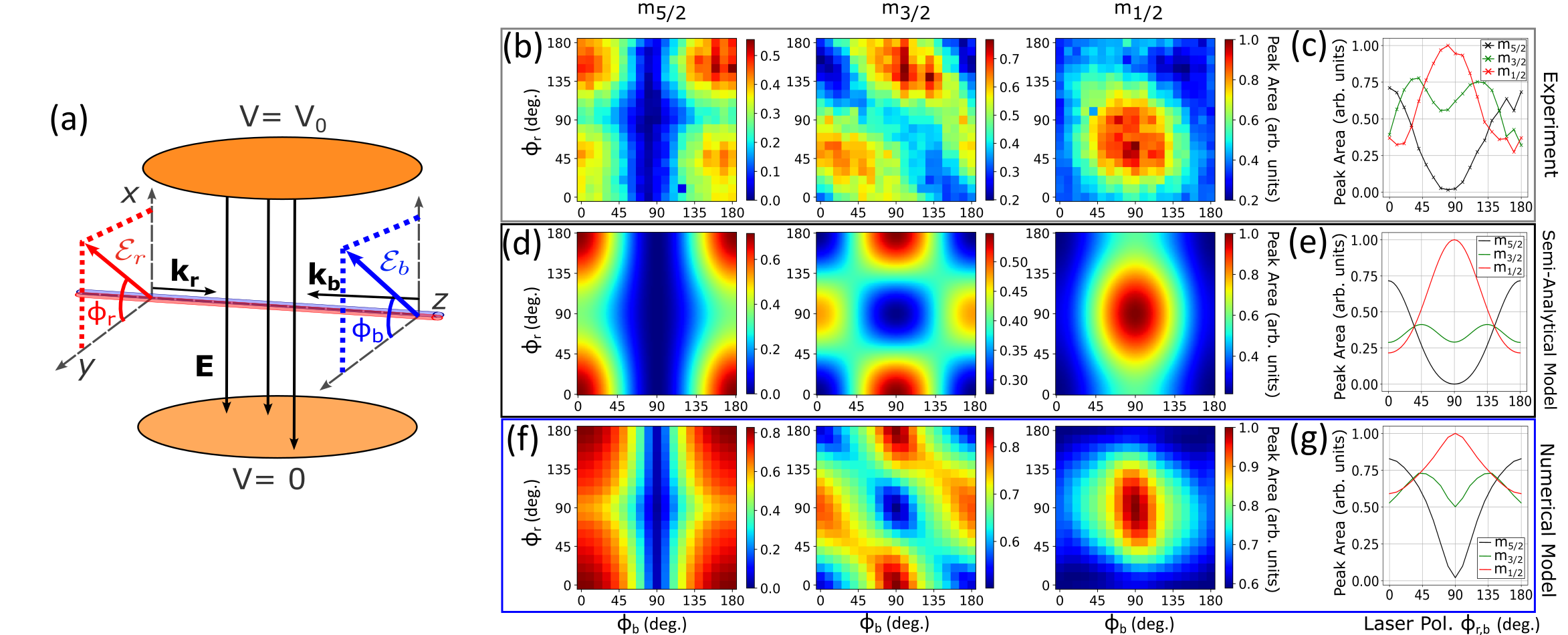}
    \caption{(a) Experimental arrangement for uniform electric field measurements. The electric field \textbf{E} points along the $x$-direction when a voltage $V_0$ is applied to the top capacitor plate. The laser beams counter-propagate along the $z$ direction, and their polarization orientations are defined by angles $\phi_r$ and $\phi_b$, formed between the laser polarization vectors $\mathcal{E}_r$ and $\mathcal{E}_b$ with the $\hat{y}$ axis. (b) Experimentally measured peak areas of $|m_J|$ EIT peaks as functions of independently varying laser polarizations. 
    (c) Experimentally measured dependence of all three peak areas on the angle between the electric field and the laser polarization when $\mathcal{E}_r$ and $\mathcal{E}_b$ are matched, i.e. $\phi_r = \phi_b$ ($\phi_{r,b}$). (d-g) Corresponding theoretical EIT area values calculated using the semi-analytical atomic model (d,e) or by solving exact numerical interaction Hamiltonian (f,g). 
    }
    \label{fig:phiRotation}
\end{figure*}


\section{Transverse uniform electric field measurements}
Our experimental setup, Fig.~\ref{fig:phiRotation}(a), consists of a vacuum chamber filled with Rb vapor at room temperature that also contains built-in capacitor plates (24~mm separation) to produce a nearly homogeneous electric field in the $x$-direction. 
To test the effect of polarization on a constant field, the bottom plate is grounded and the top plate is held at a constant $V_0 = 5$~V.
A 780~nm probe (500~$\mu\mathrm{W}$ power and $\approx 0.3~\mathrm{mm}$ wide)  laser drives the $5S_{1/2}, F=3$ ground state of ${}^{85}$Rb to an intermediate $5P_{3/2}, F=4$ excited state, while a 480~nm coupling ($50~\mathrm{mW}$ power and $\approx 0.3~\mathrm{mm}$ wide) laser drives the transition from the intermediate state to a Rydberg state. 
To calibrate the frequency axis we use the known splitting between the $46D_{5/2}$ and $46D_{3/2}$ Rydberg states as calculated using the Alkali Rydberg calculator (ARC)~\cite{ARC} and match it with the spectra measured within a reference Rb cell.
The probe and coupling laser fields counter-propagate through the vacuum chamber and the reference cell, partially canceling the Doppler shift of each atom~\cite{Finkelstein_2023,Su2024}. 
We employ two complementary measurement approaches to detect EIT resonances (and corresponding Stark shifts of the $D_{5/2}$ Rydberg energy levels) as we scan the frequency of the coupling laser. First, we measure the 780~nm probe laser transmission and identify the centroid frequencies and areas under the EIT peaks. This method provides total contribution of all Rb atoms along the laser beam and is better suited for measuring the uniform electric field produced by the capacitor plates. Second, we record the changes in the 780~nm fluorescence along the probe laser beam to obtain the local electric field information.

Fig.~\ref{fig:phiRotation}(b) experimentally demonstrates the dependence of the three $J=5/2$ EIT peaks' strengths on the polarization orientations of both red ($\phi_r$) and blue ($\phi_b$) lasers. These measurements report EIT peak areas rather than peak amplitudes. For simplicity, we use the product of the amplitude and the full-width at half maximum proportional to the peak area, using the values obtained from a Gaussian fit. This helps mitigate broadening due to local electric field gradients. 
The general peak behavior is consistent with the expectations for the semi-analytic interaction model [Fig.\ref{fig:ladder-scheme}(b)]: the $m_{\pm 5/2}$ resonance is highest for $\phi_b$ and $\phi_r$ equal to either $0^\circ$ or $180^\circ$ (aligned with $\hat{y}$-axis) when the laser polarizations are perpendicular to the electric field in the $\hat{x}$-direction, and they disappear for $\phi_b\approx\phi_r\approx 90^\circ$ when the laser polarizations and electric field vectors are aligned. The $m_{\pm 1/2}$ resonance displays the opposite behavior, increasing when the two laser polarizations are parallel to the electric field, and reducing (but not disappearing) when they are orthogonal. However, the interaction model that only accounts for the fine structure is too crude to accurately predict quantitative variations of these resonances, and fails in even qualitative prediction for the $m_{\pm 3/2}$ peak. 

\section{Theoretical models}
A more accurate accounting of the hyperfine structure is required, but exact numerical calculations accounting for the full Zeeman structure of all three electronic levels requires a lot of computational resources. Instead, we developed a simplified semi-analytical model in which the transitions to $m_J$ sub-levels of the Rydberg state are based on probabilities of transition dipole moments between the hyperfine $m_F$ states of the ground $5S_{1/2}$ and intermediate excited $5P_{3/2}$ states to the fine-structure $m_J$ states of the relevant $46D_{5/2}$ Rydberg transitions shown in Fig.~\ref{fig:ladder-scheme}(b). This model is inspired by previous studies of Zeeman-resolved Autler-Townes measurements of Rydberg states~\cite{RydbergMagnetic}. 

The EIT spectra $S(\Delta_C)$ are modeled as a sum of Gaussian peaks corresponding to the various $|m_J|$ Rydberg transitions
\begin{equation}
\begin{split}
    S(\Delta_C) = &\sum_{m_{F1}}\sum_{m_{F2}}\sum_{m_{j3}}d^2_{m_{F1}\rightarrow m_{F2}}d^2_{m_{F2}\rightarrow m_{j3}}\\ &\times \exp\left(-\frac{[\Delta_C - \Delta f_{|m_J|}(E)]^2}{2\gamma_{EIT}^2}\right),
\end{split}
\end{equation}
where $\Delta_C$ is the frequency detuning of the blue laser, $\Delta f_{|m_J|}(E)$ is the Stark shifted frequency of each $|m_J|$ peak interpolated from a Stark map numerically solved using ARC~\cite{ARC}, $\gamma_{EIT}$ is the width of the EIT resonance, $d_{m_{F1}\rightarrow m_{F2}}$ is the transition dipole matrix element between the hyperfine states of the ground and intermediate excited states, and $d_{m_{F2}\rightarrow m_{J3}}$ is the transition dipole matrix element between the hyperfine states of the intermediate excited and the fine-structure states of the Rydberg states. This approach accurately accounts for the strength of various atomic transitions for any directions of polarization and electric field vectors, but treats any possible ladder scheme independently and does not account for their interactions.

\begin{figure}
    \centering
    \includegraphics[width=1\columnwidth]{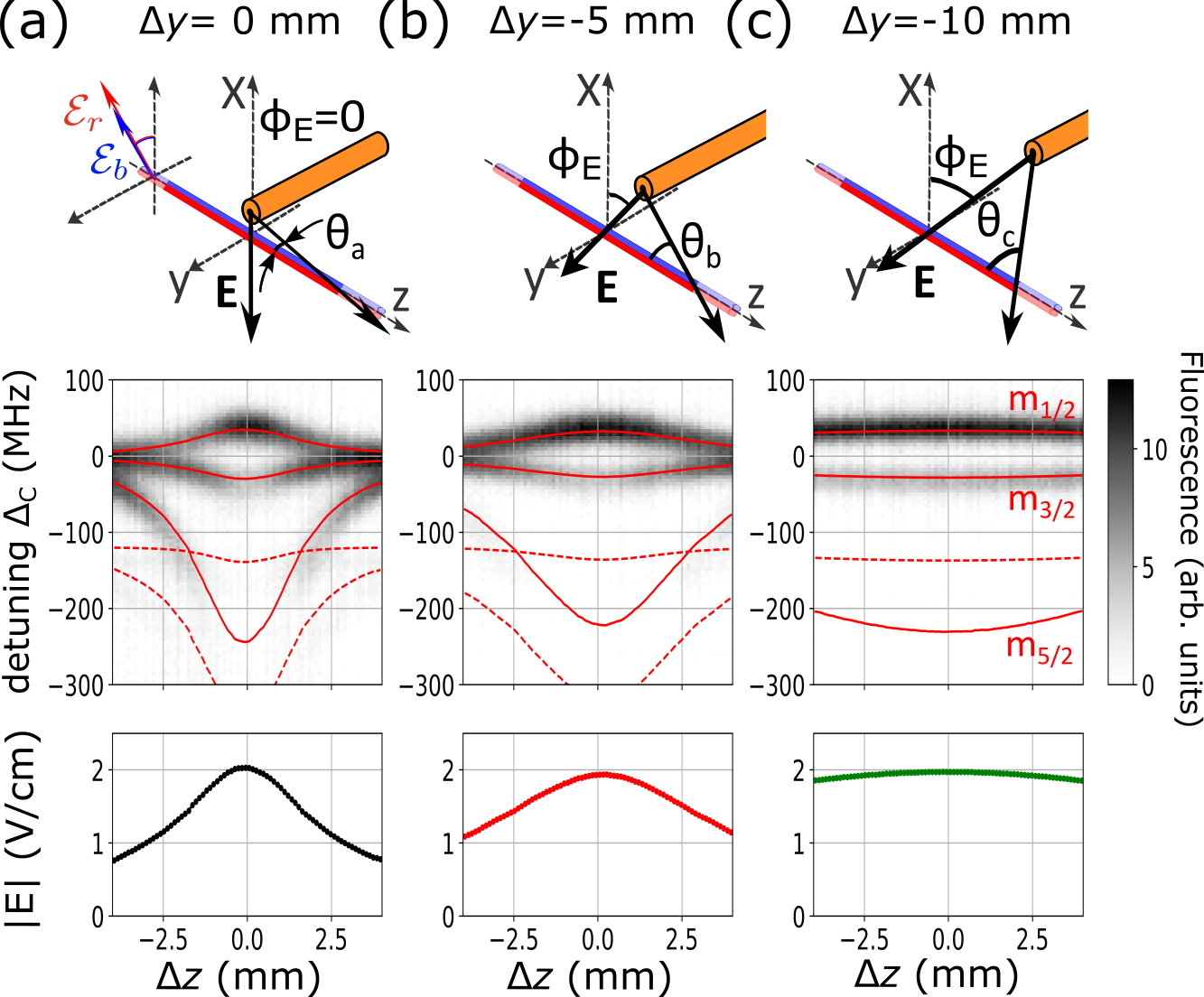}
    \caption{Fluorescence based electric field magnitude measurements. 
    (a-c) Electric field magnitude reconstruction as the wire is displaced a distance $\Delta y$ from the lasers.
    A retractable wire in our Rb vacuum chamber creates a spatially varying electric field when a voltage $V_0$ is applied. In the diagram of the wire with respect to the lasers, the darker red and blue lines indicate the laser fluorescence region monitored by the camera, $\theta_E = \theta_{a,b,c}$ denote the electric field variation with respect to the $z$-axis in each position of the probe and $\phi_E$ indicate angles with respect to the $x$- axis. 
    Underneath these diagrams, the black and white color maps are recorded fluorescence spectra for laser polarization orientation where the $m_{\pm5/2}$ peak is minimized.
    Solid lines show reconstructed peak positions of the 46D${}_{5/2}$ $|m_j|$ = 5/2, 3/2 and 1/2 and dashed red lines are the peak positions of the 46D${}_{3/2}$ levels.
    Finally, at the bottom we show the spatially reconstructed electric field magnitude for each position of the wire with respect to the lasers.
    }
    \label{fig:fluorescenceMagnitude}
\end{figure}

The dipole matrix elements, $d_{m_{Fi}\rightarrow m_{Ff}}$, depend on selection rules set by the quantization axis of the applied electric field. To account for the contribution of all laser polarizations to a specific atomic transition, we construct a ``total dipole moment'' for each transition by accounting for all allowed transitions. For the probe field we fully account for the hyperfine structure of the ground and intermediate states: 
\begin{eqnarray}
    \label{eq:dipoleMoment}
    &&d_{m_{F1}\rightarrow m_{F2}} = \sum_{\Delta m} w(\theta_E,\phi_E)_{\Delta m} \times \\&&\langle n_1,L_1,J_1,F_1,m_{F1} + \Delta m |-er |n_2,L_2,J_2,F_2,m_{F2}\rangle,\nonumber
\end{eqnarray}\label{eq:dm}  
where for each state $n$ is the principle quantum number, $L$ is the total orbital angular momentum, $J$ is the total angular momentum, $F$ is the total atomic angular momentum, $m_F$ is the azimuthal component of the total angular momentum, and $\Delta m=0,\pm1$ is the change in the $m_F$ quantum numbers based on the laser polarization.
To calculate the weights of the $\Delta m = 0, \pm 1$ transitions $w(\theta_E,\phi_E)_{\Delta m}$ we set the projection of the electric field vector onto the laser polarization axis for a given orientation of the electric field.
Examples of the angles for the electric field in the Cartesian coordinate system we use for both the probe laser polarization and electric field vectors in the lab frame are shown in Fig.~\ref{fig:fluorescenceMagnitude}. In this system the laser polarizations are rotated in the $x-y$ plane while the external electric field is allowed to be in any direction. To calculate the weights of different transitions however, we must first rotate the coordinate system such that the electric field vector sets the quantization axis 
and calculate the components of the probe laser polarization vector in this new system.

For the dipole moments $d_{m_{F2}\rightarrow m_{J3}}$ for the transitions between the intermediate and Rydberg states, we use the same approach for calculating the weights. However there is an extra complication of connecting the hyperfine structure $m_{F2}$ states to the fine structure $m_{J3}$ states. We again follow~\cite{RydbergMagnetic} and use ARC to perform the sum of the dipole matrix elements:

\begin{equation}
    \label{dipole_matrix_element}
    \small
    \begin{split}
        d_{m_{F2}\rightarrow m_{J3}} &\equiv 
        \sum_F' \sum_{m_F' = F'}^{F'} w(\theta_E,\phi_E)_{\Delta m}\sum_{m_I' = -I}^I C_{Im_{I'}J_3m_{J3}}^{F'm_{F}'}\times\\ &  \langle n_2,L_2,J_2,F_2,m_{F2}  |-er |n_3,L_3,J_3,F',m_{F}'\rangle,
    \end{split}
\end{equation}
where $C_{Im_{I'}j_3m_{j3}}^{F'm_{F}'}$ are Clebsch-Gordan coefficients and all the quantum numbers are the same as those that appear in Eq.~\ref{eq:dipoleMoment} with the addition of $I$ and $m_J$ being the total nuclear spin angular momentum and its azimuthal component, respectively. 

Fig.~\ref{fig:phiRotation}(d) shows the predictions of the semi-analytical atomic model for each EIT peak for the spectrum of probe and coupling laser polarization directions. We observe a clear similarity between these calculations and the experimental data, especially for the $m_{\pm5/2}$ and $m_{\pm1/2}$ measurements. However, our atomic model does not fully capture the behavior of the $m_{\pm3/2}$ peak, as it has the most complex resonance structure. 
Thus, in our analysis we disregard measurements of this resonance.

We also carry out exact numerical calculations of the optical transmission using a density matrix approach with the full interaction Hamiltonian and the complete hyperfine structure of all three atomic levels (details in  Appendix~\ref{app:exactmodel}). The resulting heights of Stark-shifted EIT resonances, shown in Fig.~\ref{fig:phiRotation}(f), match well with the experimental results, particularly in reproducing the proper angular dependence for the $m_{\pm3/2}$ peak. For the other two resonances the predictions of the exact numerical calculations and semi-analytical theoretical models are in reasonable agreement, thus validating the prediction of the latter. 
Because the exact numerical calculations are computationally expensive (owing to the large manifold of atomic states and Doppler broadening), this agreement gives us additional confidence in using the semi-analytical model in further analysis.

The EIT ``maps'' in Fig.~\ref{fig:phiRotation} suggest that to determine the direction of the electric field, it makes sense to align two laser polarizations and rotate them together.  
Figs.~\ref{fig:phiRotation}(c,e,g) show  variations in EIT strengths  for all three resonances for parallel polarizations: experimentally measured (c), calculated using our semi-analytical model (e) and exact numerical model (g). It is clear that such an arrangement provides maximal change in contrast in signals for the $m_{\pm1/2}$ and $m_{\pm5/2}$ EIT peaks as the laser polarizations are rotated. It is also convenient that the polarization dependencies of the $m_{\pm 5/2}$ and $m_{\pm 1/2}$ peak areas are opposite to each other, e.g., the maximum of one corresponds to the minimum of the other and vice versa. This provides a robust tool to identify the direction of the azimuthal component of the electric field vector by finding the polarization orientation corresponding to the extrema in the EIT peak strengths. 
Importantly, since we observe very good agreement between results from experimental data and both the semi-analytical and full numerical models, we can, with a fair degree of confidence, use the semi-analytical EIT model for analysis of the $m_{\pm 5/2}$ and $m_{\pm 1/2}$ EIT peaks' polarization dependence.

\section{Longitudinally varying electric field measurements}
To test our method in a more realistic situation, we apply it to characterize a spatially inhomogeneous electric field generated by a piece of thin wire (1.4~cm length and 0.8~mm diameter) placed inside the Rb chamber. Applying a voltage $V_0$ to the wire induces electric charge accumulation at the rim of the wire cylinder (confirmed by numerical solution of the Poisson equation for our experiment geometry). For simplicity we can neglect the wire thickness and approximate the charge distribution as a point charge $q_{0}$. In this case the electric field at any point is directed along the line connecting the tip of the wire and the observation point $\Delta \vec{r}=(d_x, \Delta y, \Delta z)$:
\begin{equation}
    \label{eq:wirefield}
    \vec{E}=-\frac{1}{4\pi \epsilon_0}\frac{q_{0}}{d_x^2+ \Delta y^2+ \Delta z^2}\Delta \hat{r},
\end{equation}
where $\epsilon_0$ is the vacuum permittivity constant, and $d_x=2$~mm is the vertical displacement of the laser beam from the wire. 



Capturing spatial variations in the electric field vector requires fluorescence-based EIT detection~\cite{SchlossbergerOL25,BeharyPRR2025}, where we image the fluorescence of the probe transition along a section of the laser beams near the wire and record EIT spectra for each camera pixel. We then fit these fluorescence data using the Stark shift information from the ARC database~\cite{SchlossbergerOL25,BeharyPRR2025} for each $z$-position to get the magnitude of the spatially varying electric field. Fig.~\ref{fig:fluorescenceMagnitude} shows an example of EIT fluorescence spectra and the reconstructed electric field strength at three positions of the wire: when the laser beams are directly below the wire tip $\Delta y=0$ (a), when the wire is retracted  by $\Delta y=5$~mm (b) and $\Delta y=10$~mm (c). To achieve roughly the same maximum electric field magnitude between the measurements,  we adjust the applied voltage $V_0$ for each position. 
As expected we see the strongest electric field magnitude variation along the laser beams ($z$ direction) near the wire tip, where it rapidly falls off with distance $\Delta z$. As the wire is pulled farther away we observe a smaller change in the electric field over the imaging region.

Since the laser beams have finite size, we integrate the the recorded signal across their transverse cross-section (along $x$ and $y$ dimensions of the interaction volume). If the gradient of the electric field across this region is sufficiently strong, it causes inhomogeneous broadening of EIT resonances. This broadening is particularly noticeable for the $m_{\pm5/2}$ resonance that has the largest polarizability and, therefore, is the most affected by the field gradients. As a result, for some of the following analysis we mostly rely on the $m_{\pm1/2}$ resonance measurements. We note, however, that the $m_{\pm5/2}$ resonance becomes more attractive for Rydberg EIT measurements  at lower $n$ Rydberg states~\cite{vorobiov2025arxiv}.

 \begin{figure}[ht]
    \centering
    \includegraphics[width=1\columnwidth]{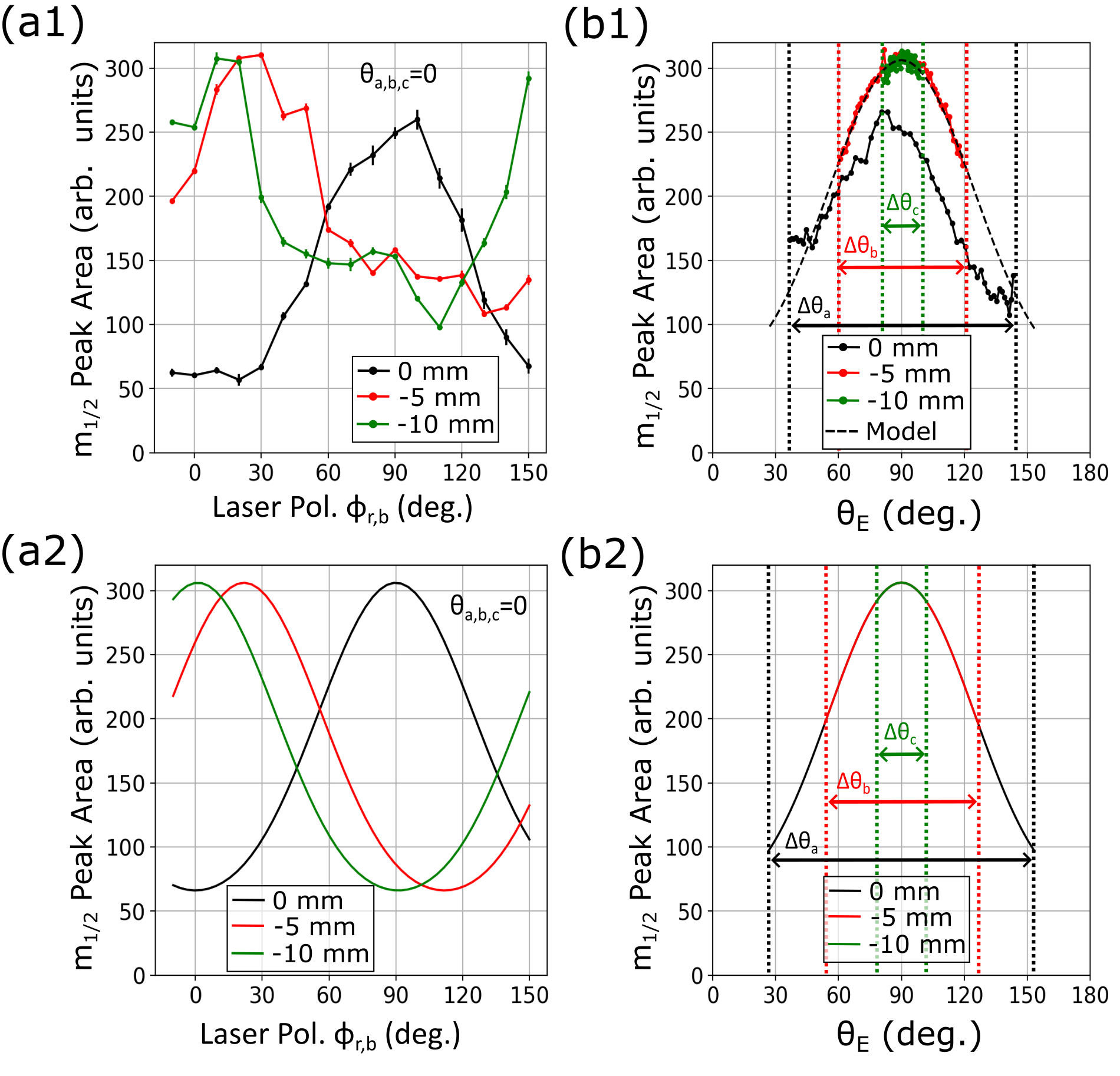}
    \caption{Variation of $\theta_E$ and $\phi_E$ for different $\Delta y$ wire displacements.
    The different $\Delta \theta_{a,b,c}$ correspond to the different angular variations we show in Fig.~\ref{fig:fluorescenceMagnitude}.
    (a1) Experimental $m_{\pm1/2}$ peak area for different wire $\Delta y$-positions, monitored with fluorescence at $\Delta z$ = 0 mm. Error bars are an average of pixels from the center $\pm0.25$~mm. 
    (b1) Experimental $m_{\pm1/2}$ peak area for different wire $\Delta y$-positions with a fixed polarization when $m_{\pm5/2}$ peak is minimum.
    (a2) Modeled $m_{\pm1/2}$ peak area for different wire $\Delta y$-positions.
    (b2) Modeled $m_{\pm1/2}$ peak area for different wire $\Delta y$-positions with a fixed polarization when the $m_{\pm5/2}$ peak is at a minimum.
    }
    \label{fig:thetaPhiVar}
\end{figure} 

To characterize the electric field direction we use two angles: longitudinal $\theta_E$ (defined as an angle between the electric field and the laser propagation direction), and azimuthal $\phi_E$ (defined as an angle between the transverse $x-y$ component of the electric field and $x$-axis), as shown in the diagrams of the field generated by the wire in Fig.\ref{fig:fluorescenceMagnitude}. 
For example, in the earlier measurements the vertical electric field was perpendicular to the laser beam propagation: $\theta_E=90^\circ$ and $\phi_E=0^\circ$. Using Eq.(\ref{eq:wirefield}), we can estimate values for both angles:
\begin{eqnarray}
   \cos (\phi_E) &=&  \frac{\Delta  y}{\sqrt{d_x^2+\Delta y^{2}}} \label{eq:phi}
,\\
    \cos(\theta_E) &=& \frac{\Delta  z}{\sqrt{d_x^2+\Delta y^{2}+\Delta  z^{2}}}.\label{eq:theta}
\end{eqnarray}

The value of the azimuthal angle $\phi_E$ depends primarily on the distance between the wire and the laser beam, and does not depend on the position along the laser beam. In the experiment we can find the azimuthal angle $\phi_E$ at each location by synchronously rotating the laser polarizations $\mathcal{E}_{r,b}$ and finding the angle at which the amplitude of the $m_{\pm1/2}$ fluorescence peak is maximized. 
Fig.~\ref{fig:thetaPhiVar}(a1) shows examples of such measured dependence for three different distances $\Delta y$ of the wire away from the laser beam, measured at the center $\Delta z = 0$~mm. 
We observe that when the wire is directly above the lasers ($\Delta y = 0$~mm), the area of the $m_{\pm1/2}$ EIT peak is maximized when the laser polarizations are nearly vertical ($\phi_b=\phi_r = 90^{\circ}$). 
As the wire is retracted, this maximum shifts toward horizontal polarizations ($\phi_b=\phi_r \rightarrow 0^{\circ}$), indicating the expected change in $\phi_E$. For comparison, in Fig.~\ref{fig:thetaPhiVar}(a2) we plot the $m_{1/2}$ amplitudes calculated using our semi-analytic model and assuming an electric field distribution given by Eq.(\ref{eq:wirefield}), which then agree well with experiment.


When we track the change in electric field direction along the laser beam, it primarily occurs in the longitudinal direction, as given by Eq.(\ref{eq:theta}).
Note that this variation depends on the horizontal distance $\Delta y$ between the wire and the laser beam and we investigate three different wire positions in Fig.\ref{fig:fluorescenceMagnitude}.
Since we monitor the same absolute $z$-range for each wire position, we expect to sample a smaller $\theta_E$ variation when the wire is farther away from the lasers.
At the same time, measurement of $\theta_E$ is more challenging since the longitudinal component of the electric field does not fundamentally change the interaction symmetry and, consequentially, the polarization dependence of the EIT peaks. Instead, reducing the angle between the electric field and the laser beams makes the polarization dependence less pronounced, reducing the span of azimuthal variations and eventually eliminating any polarization dependence as $\theta_E \rightarrow 0^\circ$. For our experimental arrangement we thus expect to find the $m_{\pm1/2}$ EIT peak to be the strongest for $\theta_E=90^\circ$ and then become smaller as $\theta_E$ decreases. 

Fig.\ref{fig:thetaPhiVar}(b1) shows the experimental dependence of the $m_{\pm1/2}$ EIT peak area along the laser beam on the angle $\theta_E$, calculated using Eq.(\ref{eq:theta}). For each wire position, we adjust the laser polarizations to maximize the EIT contrast. The corresponding semi-analytic model calculations are shown in Fig.\ref{fig:thetaPhiVar}(b2). As expected, we observe that the experimental EIT data generally follow the expected behavior: the measured resonance peak areas are largest at the position corresponding $\theta_E\approx90^\circ$, and they all have similar shapes but cover different range of angles. Specifically, the data for $\Delta y=-5$~mm and $\Delta y=-10$~mm positions match the semi-analytic theory predictions extremely well. There is some deviation in overall resonance strength for $\Delta y=0$ (most likely caused by some known imperfection in the wire and the effect of stronger field gradients, or our assumption that the wire acts as a point charge breaks down this close to the wire), but it still closely follows the expected dependence. This gives us some confidence that with better calibration it should be possible to use the EIT resonance analysis to reconstruct full vector information of inhomogeneous electric fields. 

It is important to note that our method intrinsically does not allow us to distinguish between $\phi_E$ and $\phi_E+180^\circ$ as well as between $\theta_E$ and $180^\circ - \theta_E$, as these angles result in identical interaction schemes. 
To uniquely identify the electric field direction we may need another degree of freedom. 
For example, introducing an additional magnetic field to set a clear quantization axis~\cite{RydbergMagnetic,chen2025polarizationawaredoadetectionrelying}. However, this approach requires creating a more robust model to describe the interplay of Stark and Zeeman effects and will be a subject of future studies. 
Another challenge is that in our current experimental setup we cannot directly vary the orientation of the electric field, and thus it is impossible to experimentally verify the effect $\theta_E$ has on EIT resonances independently. Instead, we have to rely on the semi-analytical model for this information. 


\section{Conclusion}
In conclusion, we demonstrated that the strong dependence of various Stark-split $m_J$ EIT resonances on the polarizations of optical fields can be useful for vector electric field measurements. Moreover, we showed that these changes are accurately described by a semi-analytic atomic model for the $m_{\pm5/2}$ and $m_{\pm1/2}$ Rydberg states, but a more complete model and additional experimental verification are needed to describe the full EIT spectrum more accurately. The reported results suggest a viable approach for vector electric field measurements that will enable accurate reconstruction of electric charge distributions for, e.g., electron beam characterization~\cite{BeharyPRR2025} and plasma diagnostics~\cite{PhysRevE.63.047401,vorobiov2025arxiv,dash2025electricfielddiagnosticscontinuous}.

\section*{Acknowledgments}
This work is supported by U.S. DOE Contract DE-SC0024621 and DE-AC05-06OR23177, NSF award 2326736 and Jefferson Lab LDRD program.

\appendix

\section{Exact numerical calculations details}\label{app:exactmodel}

The Rydberg system can be modeled by a atomic energy level structure with 3 state manifolds (ground state, excited state and Rydberg state). Each manifold contains sublevels that are affected differently and are interconnected in a complex web-like graph. The overall dynamics of the system are contained in its time-dependent density matrix ${\boldsymbol{\rho}}(t)$. 

System evolution due to all the processes is governed by the Liouville-Von Neumann equation, given by
\begin{equation}\label{Eq:Master}
	\frac{d{\boldsymbol{\rho}}(t)}{dt} = -\frac{i}{\hbar} [{\boldsymbol{H}},{\boldsymbol{\rho}}(t)] + {\boldsymbol{L}}.
\end{equation}

In the interaction picture, the time-evolution of the density matrix is determined by a Hamiltonian matrix operator ${\boldsymbol{H}}$ containing atom-electromagnetic field interactions. In a 2-photon Rydberg system, the coupling strengths between each pair of energy levels are written in terms of the amplitude and detunings of the laser fields in a laser Hamiltonian term $\boldsymbol{H}_{laser}$, and a separate dc electric field-dependent Stark Hamiltonian matrix $\boldsymbol{H}_{Stark}$ that only affects their energy shifts, $\Delta_{st}(E_{DC})$. 

\begin{equation}
	\boldsymbol{H}=\boldsymbol{H}_{laser}(E_p,\Delta_p,E_c,\Delta_c)+\boldsymbol{H}_{Stark}(E_{DC})
	\label{eq:ham1}
\end{equation}


We can assign a label ``$i$" to each sublevel in a manifold (indexed g, e, and r for ground, excited, and Rydberg, respectively). Since each sublevel belongs to its own independent manifold, we can write the Hamiltonian operator as a block matrix. Here we will assume that the Stark shifts in the ground and excited state manifolds are negligible due to their much smaller polarizabilities compared to the Rydberg state manifold and ignore their contribution to the Stark matrix.

\begin{equation}
	\boldsymbol{H}=
	\begin{pmatrix}
		\boldsymbol{H}_{gg} & \boldsymbol{H}_{ge} & 0 \\
		\boldsymbol{H}_{eg} & \boldsymbol{H}_{ee} & \boldsymbol{H}_{er} \\
		0 & \boldsymbol{H}_{re} & \boldsymbol{\Delta}_{rr} 
	\end{pmatrix}.
	\label{eq:ham}
\end{equation}

Each submatrix along the diagonal of $\boldsymbol{H}$ is itself a diagonal matrix that only contains the laser detuning or Stark shift of the sublevel in that manifold (i.e. $\Delta_{ii}$). The off-diagonal submatrices are rather complicated; the coupling strength between sublevels is not only dependent on the laser intensity. Each term is scaled by the transition's static dipole matrix element 
but also the laser polarization angle-dependent weight of that optical transition. Within each submatrix $\boldsymbol{H}_{IJ}$, the terms $\boldsymbol{H}_{ij}$ can be written as

\begin{equation}
	\boldsymbol{H}_{ij} = \frac{\hbar}{2}\vec{d}_{ij}\cdot \vec{E} 
\end{equation}
where $\left|\vec{E}\right|$ is the laser field strength and 
\begin{equation}
	\left|\vec{d}_{ij}\right|=
    \sum_{q} A_q \mathcal{C}_{ij;q} \left<i\left|-e\textbf{r}\right|j\right>
\end{equation}
is a decomposition of the coupling strength in terms of the dipole matrix element between the ground/excited/Rydberg states $\left<i\left|-e\textbf{r}\right|j\right>$ and the associated Clebsch-Gordon coefficient $\mathcal{C}_{ij;q}$ that connects sublevels $i$ and $j$ for a spherical tensor index $q$ for each type of optical transition (i.e. $q= -1,0,1 $ for $\sigma_{-}$, $\pi$, $\sigma_{+}$ transitions, respectively.). Meanwhile, the polarization angle-dependent weight $A_q$ for linearly polarized fields is written in terms of the angle $\theta$ with respect to the lab frame vertical direction:

\begin{equation}
	\begin{pmatrix}
		A_{-}  \\
		A_0 \\
		A_{+}  
	\end{pmatrix} = 
	\begin{pmatrix}
		\frac{1}{\sqrt{2}}\text{sin}(\theta)  \\
		\text{cos}(\theta) \\
		\frac{-1}{\sqrt{2}}\text{sin}(\theta)  
	\end{pmatrix}
	\label{eq:ham2}
\end{equation}

In addition to the application of external fields, environmental interactions also contribute to the atomic system evolution. These are captured in the Lindblad operator ${\boldsymbol{L}}$, where each decay or decoherence mechanism has its own term. In the model we consider the natural decay of each excited state ($\Gamma_{21}$, $\Gamma_{32}$), as well as dephasing associated with atoms transiting in and out of the laser fields. Using a parameter $\gamma_t$ we quantify the transit dephasing as the rate at which atoms in the Rydberg state depopulate back down to the ground state. The Lindblad operator matrix is therefore defined as

\begin{equation}
	\begin{split}
		\boldsymbol{L}_{total}=\Gamma_{eg}\boldsymbol{L}_{eg}+\Gamma_{re}\boldsymbol{L}_{re}+
		+\gamma_{t}\boldsymbol{L}_{rg}.
	\end{split}
	\label{eq:lindblad1}
\end{equation}

Within a submatrix $\boldsymbol{L}_{IJ}$ for a manifold pair $I,J$ underlies a contribution $\boldmath{L}_{ij}$ from each sublevel, that is defined in terms of jump operators as

\begin{equation}
	\Gamma_{ij}\boldsymbol{L}_{ij} = \frac{\Gamma_{ij}}{2}(2\boldsymbol{\sigma}_{ji}\boldsymbol{\rho}\boldsymbol{\sigma}_{ij} - \boldsymbol{\sigma}_{ii}\boldsymbol{\rho} - \boldsymbol{\rho}\boldsymbol{\sigma}_{ii})
\end{equation}
Each quantity $\Gamma_{ij}$ also needs to be weighed appropriately by factors of $\vec{d}_{ij}$ so as to ensure that the total depopulation rate from a sublevel $i$ adds up to unity.


Calculating the probe laser transmission through the atomic vapor requires knowledge of $\rho_{ge}$, whose imaginary part determines the effective absorption coefficient $\chi$ for the probe field as it passes through the atomic vapor: 
$\chi \propto \text{Im}(\rho_{12})$. We only consider the case where the system is in steady-state so that $\boldsymbol{\rho}(t) = \boldsymbol{\rho}_0$. Therefore Eq.~\ref{Eq:Master} reduces to

\begin{equation}
	0 = -\frac{i}{\hbar} [{\boldsymbol{H}},{\boldsymbol{\rho}}] + {\boldsymbol{L}}.
\end{equation}
and can be solved for $\boldsymbol{\rho}_0$ which depends on all the parameters that are tuned during the experiment.

The analysis above is performed for one set of probe and control field detunings, $\Delta_p$ and $\Delta_c$, which are contained in $\boldsymbol{H}$. To include finite temperature effects where the atoms are moving around in the vapor cell, Doppler-induced laser detunings must be taken into account. We add these as Doppler shifts from velocities along the optical field propagation that are sampled from a thermal Maxwell distribution according to
\begin{equation}
	f(u)=\frac{1}{\sqrt{\pi}}\int_{-\infty}^{\infty}e^{-u^2}du,
	\label{eqn:maxwell}
\end{equation}
where $u\equiv{v/\sigma_v}$, $\sigma_v=\sqrt{2 k_B T/m}$, $v$ is the atom speed, $k_B$ is Boltzmann's constant, $T$ is the average atomic temperature, $m$ is the atomic mass. The effective detunings of the probe and control lasers for any velocity class $u$, set by the root-mean-square velocity for the thermal distribution, are given by the following:
\begin{equation}
	\begin{array}{c}
		\Delta_{p}^{'}(u)=\Delta_{p,0}+\frac{2\pi\sigma_v}{\lambda_{p}}u \\
		\\
		\Delta_{c}^{'}(u)=\Delta_{c,0}-\frac{2\pi\sigma_v}{\lambda_{c}}u.
	\end{array}
	\label{eqn:effdets}
\end{equation}
Here, $\Delta_{p,0}$ and $\Delta_{c,0}$ are the overall detunings of the probe and control fields from the atomic resonances and $\lambda_{p}$ and $\lambda_{c}$ are their wavelengths, respectively. With this correction,
\begin{equation}
	\boldsymbol{\rho}_0 \equiv \boldsymbol{\rho}_0(\Delta_{p}^{'},\Delta_{c}^{'}) = \boldsymbol{\rho}_0(u)
\end{equation}

To obtain the Doppler-averaged density matrix, we need to numerically integrate over all velocity classes:
\begin{equation}
	\bar{\boldsymbol{\rho}_0}=\frac{1}{\sqrt{\pi}}\int_{-\infty}^{\infty}e^{-u^2}\boldsymbol{\rho}_0(u)du.
	\label{eqn:doppintfinal}
\end{equation}
This result is then used to extract $\text{Im}(\rho_{12})$, giving the absorption coefficient.

\bibliography{bibliography}

\end{document}